\DocumentMetadata{
  lang        = en-US,
  pdfstandard = ua-2,
  tagging=on
  }

\documentclass[11pt]{article}

\usepackage[a4paper,margin=1in]{geometry}

\usepackage[T1]{fontenc}
\usepackage{mathpazo}          
\usepackage{microtype}         
\usepackage{graphicx}
\usepackage{booktabs}
\usepackage{caption}
\usepackage{float}
\usepackage{titlesec}
\titleformat{\section}{\normalfont\large\bfseries}{\thesection}{1em}{}
\titleformat{\subsection}{\normalfont\normalsize\bfseries}{\thesubsection}{1em}{}
\titlespacing*{\section}{0pt}{1.4em}{0.8em}
\titlespacing*{\subsection}{0pt}{1.1em}{0.6em}

\usepackage{authblk}

\usepackage[numbers,sort&compress]{natbib}

\usepackage{url}
\usepackage{xcolor}
\definecolor{linkcolor}{RGB}{0,0,0}
\usepackage[colorlinks=true,linkcolor=linkcolor,citecolor=linkcolor,urlcolor=linkcolor]{hyperref}

\title{\bfseries Why Users Stop at Good Enough: Applying Bounded Rationality to Self-Service Restaurant Display Interactions}

\author[1]{Muhammad Ismail Khursheed}
\author[1]{Carolina Islas Sedano}
\author[1]{Jaakko Järvi}
\affil[1]{Department of Computing, University of Turku, Finland}
\date{}
\begin{document}

\maketitle

\begin{abstract}
This paper investigates user interaction with digital displays in self-service restaurant environments, focusing on the phenomenon where users often settle for good enough interactions rather than optimal. We apply bounded rationality theory to understand how cognitive limitations influence user decisions to engage with displays of varying interaction complexity. Through analysis of two field experiments conducted at a self-service restaurant in Finland, we examine interactive displays that require different levels of cognitive investment: a simple binary choice interface and a multi-step interaction. Our initial findings suggest an "engagement plateau" where increased interaction complexity yields diminishing returns in initial engagement rates, yet users who overcome initial cognitive barriers demonstrate high completion rates. This work contributes to Human-Computer Interaction field by offering an empirically supported theoretical lens for understanding user behavior in Self-Service Restaurant (SSR) environments and providing actionable design principles for public interactive displays that accommodate human cognitive limits.
\end{abstract}


\section{Introduction}
Interactive displays are increasingly deployed in public spaces, from transportation hubs to self-service restaurants (SSRs)~\cite{10.1007/978-3-642-01516-8_1, 10.1007/978-3-540-79576-6_14, 10.1145/2307798.2307815}. However, engaging users in dynamic, distracting environments remains a significant Human-Computer Interaction (HCI) challenge~\cite{10.1007/978-3-642-01516-8_1, 10.1007/978-3-540-79576-6_14}. Real-world observations consistently show users deviating from idealized interaction models, often engaging minimally with available features or abandoning interactions before completion~\cite{1586033638, 10.1109/MC.2012.115}.

In SSR environments, we have observed that users frequently exhibit a pattern of settling for good enough\footnote{Good Enough: A choice or solution that meets minimum acceptable criteria without necessarily being optimal, reflecting a pragmatic approach to decision-making under constraints.} interactions rather than optimal interactions. This behavior reflects a pragmatic adaptation to the cognitive constraints inherent in public spaces, where users must balance multiple competing demands on their attention and cognitive resources~\cite{10.1037/0033-2909.116.2.220}.

This work draws its inspiration from existing research~\cite{khursheed2025affordance} that developed an affordance-based design model for interactive screens in SSR environments. The research further investigated how varying levels of communication, passive\footnote{Passive Communication: A display mode that provides information without requiring direct user input or interaction.}, active\footnote{Active Communication: A display mode that requires explicit user action, such as tapping or clicking, to register a response or proceed.}, and two-way\footnote{Two-Way Communication: An interactive display mode that involves multiple steps of user input and system response, creating a dialogue between user and system.}, influence user engagement and actions in a SSR. While the original research focused on the design and effectiveness of these interactive systems~\cite{khursheed2025affordance}, the empirical data collected reveals a compelling and consistent pattern: users limited their engagement efforts, even when more information or potentially superior options were readily available. This observation prompted a deeper inquiry into the underlying cognitive mechanisms driving such behavior. Similar patterns of satisficing and limited engagement have also been observed in other public-facing, time-constrained self-service systems, such as ticketing kiosks, quick service restaurants, Automated Teller Machines, hotel check-ins, etc~\cite{10.1177/1094670502238916, Lee_Oh_2022}.

We interpret user engagement patterns through the theory of bounded rationality\footnote{Bounded Rationality: A theory proposing that rationality is limited when individuals make decisions, and under these limitations, rational individuals will select a decision that is satisfactory rather than optimal~\cite{10.1080/09672567.2018.1523206}.}, developed by Herbert A. Simon~\cite{10.1037/h0042769}. This theory posits that human decision-making is constrained by cognitive limitations, incomplete information, and time pressures. Rather than making globally optimal choices, individuals make decisions that are reasonable given these constraints~\cite{10.7551/mitpress/1654.001.0001}. In the context of public interactive displays, bounded rationality can help to explain why users often choose simpler interaction paths even when more comprehensive or potentially beneficial options are available~\cite{10.7551/mitpress/4702.001.0001}.

Our research addresses the following Research Question: \textbf{How does bounded rationality explain user engagement patterns with interactive displays of varying complexity in self-service restaurant environments?}

To address this question, we analyze data from two field experiments examining interactive displays with different complexity levels: a simple binary choice interface\footnote{Binary Choice Interface: A simple interactive display requiring users to select between two options with a single tap.} and a multi-step interaction\footnote{Multi-Step Interface: An interactive display requiring multiple user actions and system responses to complete the full interaction sequence.}. These experiments allow us to investigate how interaction complexity\footnote{Interaction Complexity: The cognitive effort required to complete an interaction, determined by factors such as the number of steps, decision points, information processing requirements, and motor actions needed.} influences user engagement decisions, providing insights into the cognitive mechanisms that drive users to stop at good enough rather than pursue optimal interactions.
\section{Related Work}
To map our understanding of user engagement with interactive self-service restaurant displays, we explore prior research in three domains central to our study: public display interaction, cognitive load and decision-making in HCI, and behavioral economics in interface design. This review establishes how earlier work has framed user engagement and decision-making in similar contexts, setting the stage for our subsequent theoretical framing with bounded rationality in the Theoretical Foundations section (see Section~\ref{theoretical-foundations}).

\subsection{Public Display Interaction Research}
Public interactive displays have been extensively studied in HCI literature, with researchers investigating various aspects of user engagement and interaction patterns~\cite{10.1109/MCG.2013.28, 10.1145/1873951.1874203}. Early work by Brignull and Rogers~\cite{1586033638} established foundational understanding of how to entice users to interact with large public displays, identifying key factors such as content relevance, social acceptability, and interaction simplicity.

Subsequent research has explored the "display blindness" phenomenon, where users fail to notice or engage with digital signage despite its prominent placement~\cite{10.1007/978-3-642-01516-8_1}. Müller~\cite{10.1007/978-3-642-01516-8_1} demonstrated that user expectations significantly influence attention toward digital displays. Dalton~\cite{10.1145/2702123.2702150} used eye-tracking to provide more nuanced understanding of situated display visibility.

Furthermore, Vogel and Balakrishnan~\cite{10.1145/1029632.1029656} introduced the notion of transitioning from implicit to explicit interaction, demonstrating how public displays can accommodate different levels of user engagement. This work laid the groundwork for our understanding of progressive engagement, which is reflected in the design of our second experiment involving a multi-step interface (see Section~\ref{exp-2-design}).

\subsection{Cognitive Load and Decision-Making in HCI}
Cognitive load theory, originally developed by Sweller~\cite{10.1207/s15516709cog1202_4}, has been widely applied in HCI to understand how interface complexity affects user performance~\cite{10.1016/S0747-5632(02)00086-9, brainsci15020203}. Research has shown that excessive cognitive load can lead to decreased task performance, increased errors, and user abandonment~\cite{10.1177/154193120905302206, 10.48550/arXiv.2402.11820}. Our work extends this understanding by applying bounded rationality theory to explain why users make "good enough" choices rather than optimal ones.

Decision-making research in HCI has increasingly recognized the importance of cognitive constraints and heuristics~\cite{10.1145/3706598.3713450, 10.17932/EJEAS.2021.024/ejeas_v03i2001}. Work by Payne~\cite{payne1988adaptive} on adaptive decision-making provides theoretical foundation for understanding how users adjust their decision strategies in time-constrained environments. Our research complements and aligns with the previous work on engagement plateaus and completion patterns (see Section~\ref{theoretical-foundations}).

\subsection{Behavioral Economics in Interface Design}
The application of behavioral economics principles to interface design has gained significant attention in recent years~\cite{10.1145/3290605.3300733}. Thaler and Sunstein's work on nudging~\cite{9780141040011} has influenced numerous HCI studies that explore how interface design can guide user behavior without restricting choice~\cite{weinmann2016digital, 10.1145/3290605.3300733}. Our application of bounded rationality theory contributes to this growing body of work by providing empirical evidence for how cognitive limitations manifest in public display contexts.

Furthermore, research on sunk cost effects and commitment escalation~\cite{arkes1999sunk, 379023c0-b156-3eeb-958e-4883fb3086f6, STAW197627} has shown how initial investments influence subsequent decision-making. Our findings about high completion rates among users who overcome initial barriers align with this literature and extend it to public interactive display contexts.

\section{Theoretical Foundations} \label{theoretical-foundations}
Our investigation focuses on two key phenomena that emerge from bounded rationality theory in the context of public interactive displays: engagement plateaus and commitment effects in completion patterns.

\subsection{Engagement Plateau}
Bounded rationality theory~\cite{10.1037/h0042769} proposes that human decision-making is fundamentally constrained by cognitive limitations, incomplete information, and time pressures. In public interactive display contexts, these constraints manifest as engagement plateaus\footnote{Engagement Plateau: The term 'engagement plateau' is introduced by the authors to describe a phenomenon where increasing interaction complexity does not proportionally increase user engagement rates, suggesting cognitive limits to user investment.} where increased interaction complexity does not proportionally increase user engagement rates. We identify two interrelated mechanisms: users’ limited cognitive capacity, which establishes engagement thresholds, and the cognitive costs of complexity, which shape the engagement–complexity relationship.

Cognitive Load and Engagement Thresholds: Users have finite cognitive capacity for processing information and making decisions~\cite{kahneman1973attention, 10.1145/1054972.1055101}. In SSR environments, users must simultaneously process visual information from displays, navigate physical space, make food choices, and often manage social interactions. This cognitive multitasking creates conditions where users establish implicit thresholds for the amount of mental effort they are willing to invest in display interactions~\cite{10.1145/1541948.1541999}.

Complexity-Engagement Relationship: As interaction complexity increases, requiring more steps for decisions or information processing, the cognitive cost of engagement rises. Bounded rationality theory suggests that beyond a certain threshold, additional complexity will yield diminishing returns in user engagement as the perceived cognitive cost outweighs the perceived benefits for most users~\cite{10.1080/09672567.2018.1523206, 10.7551/mitpress/1654.001.0001}.

\subsection{Commitment and Completion Effects}
Once users overcome initial cognitive barriers and decide to engage with interactive displays, different decision-making patterns emerge regarding completion behaviors.

Users who begin complex interactions have already made a deliberate decision to invest cognitive effort. This initial investment functions as a sunk cost that influences subsequent decision-making, making completion more attractive than abandonment~\cite{ARKES1985124}. From a bounded rationality perspective, once the cognitive barrier has been crossed, satisficing behavior makes interaction the \emph{good enough} choice compared to abandoning the interaction without outcome~\cite{THALER198039}. Complementary theories of sunk cost~\cite{arkes1999sunk}  and escalation of commitment~\cite{379023c0-b156-3eeb-958e-4883fb3086f6} further explain why users persist in multi-step interactions even when tasks prove more complex than anticipated. 

These theoretical foundations provide a lens for understanding why users exhibit different behaviors at different stages of interaction with public displays, and how cognitive constraints and commitment dynamics together shape both engagement initiation and completion patterns.

\section{Research Design}
This study analyzes data from field experiments originally conducted to investigate the design and effectiveness of different communication modalities on interactive screens to influence user engagement in real-world setting~\cite{khursheed2025affordance}. The study employed the Hevner’s Three Cycle View of Design Science Research~\cite{hevner2007three} methodology. Our analysis applies bounded rationality theory to understand the cognitive mechanisms underlying user engagement decisions with displays of varying interaction complexity.

\subsection{Study Setting: A Self-Service Restaurant}
All experiments were conducted at Flavoria Research Lab, a university self-service buffet restaurant and lab located in Turku, Finland. Flavoria Research Lab serves 500 to 1000 diners daily during working days, providing a naturalistic environment for studying user behavior with interactive displays. The setting offers several advantages for research: consistent traffic patterns, a regular user population of students, faculty, and staff, and stable operational procedures that allow measurement of intervention impacts.

The diverse user population encompasses varying demographics, technological literacy levels, and familiarity with the space, making it representative of many public interactive display contexts. The restaurant's commitment to sustainability initiatives provided meaningful content for experiments focused on user engagement.

Furthermore, the SSR environment presents several challenges relevant to bounded rationality research. Interactive displays must compete with numerous competing stimuli including food selection, social interaction, and personal devices. Users operate under time pressures from hunger, social expectations, and scheduling constraints. The public nature of interactions introduces social considerations that may influence engagement decisions. These factors create cognitive demands that make the setting fruitful for studying how bounded rationality affects user engagement with interactive displays.

\subsection{Background Context: Passive Information Display}

\begin{figure}
    \centering
    \includegraphics[width=0.5\textwidth, alt={Restaurant is closed: Image implying the restaurant is currently closed and will be open from 10:30 to 13:00}]{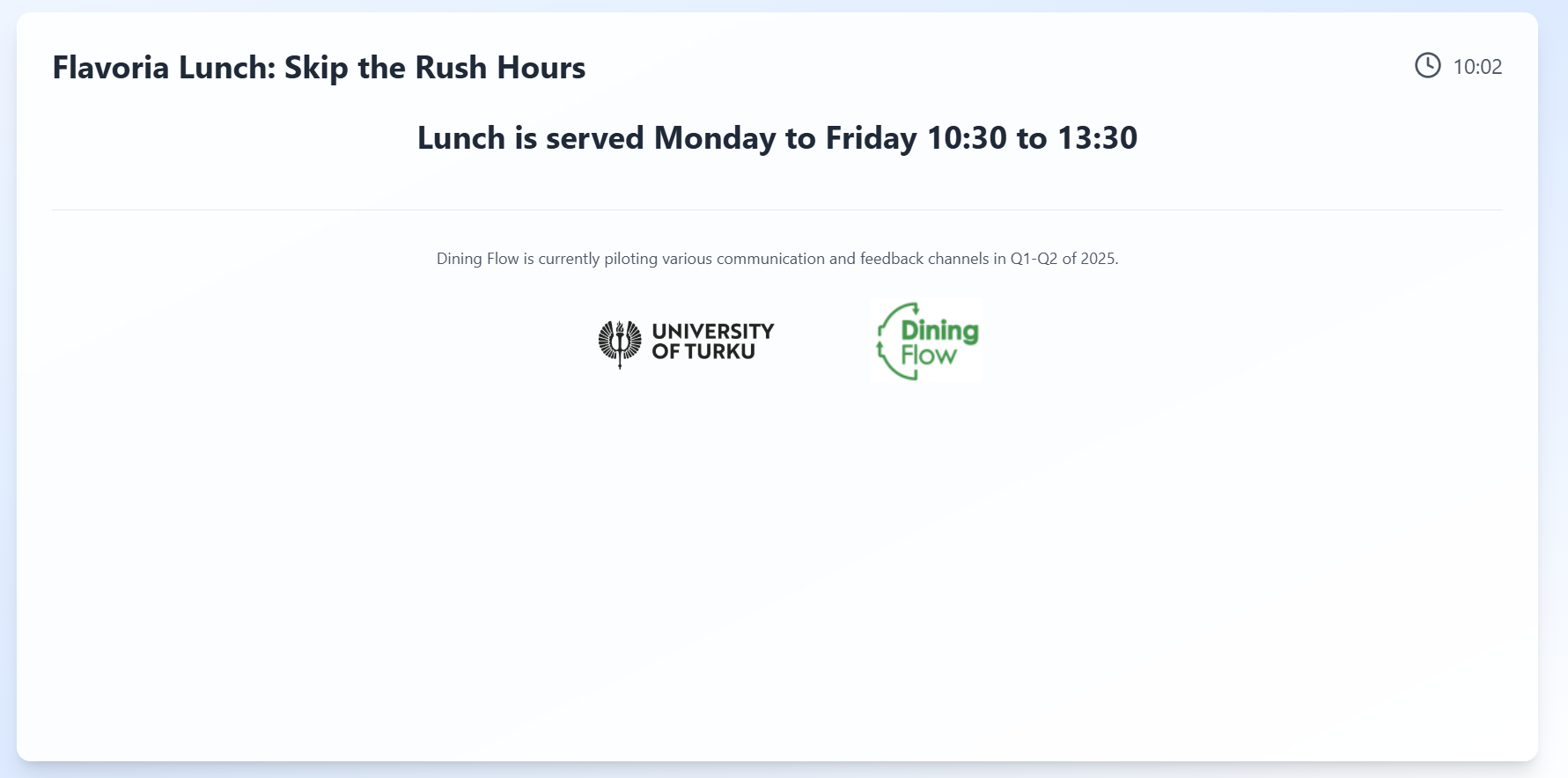}
    \caption{Passive Information Display: Closed State}
    \label{fig:passive-display-closed}
\end{figure}
\begin{figure}
    \centering
    \includegraphics[width=0.5\textwidth, alt={Wait times of 5 minutes: Image showing current wait times of 5 minutes and a yellow progress bar stating Moderate wait times}]{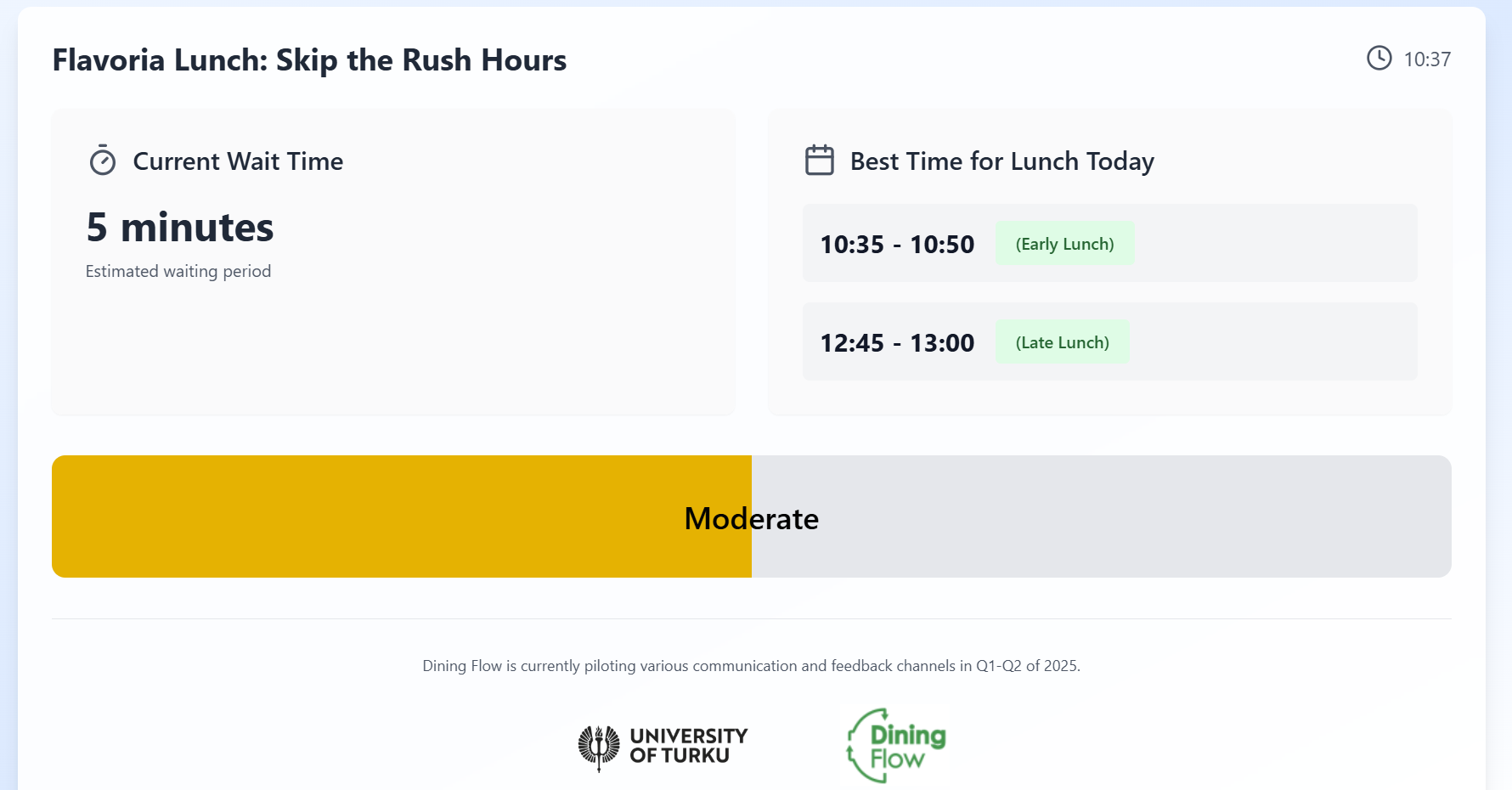}
    \caption{Passive Information Display: Moderate State}
    \label{fig:passive-display-moderate}
\end{figure}
\begin{figure}
    \centering
    \includegraphics[width=0.5\textwidth, alt={Wait times of 10+ minutes: Image showing current wait times of 10+ minutes and a red progress bar stating Busy wait times}]{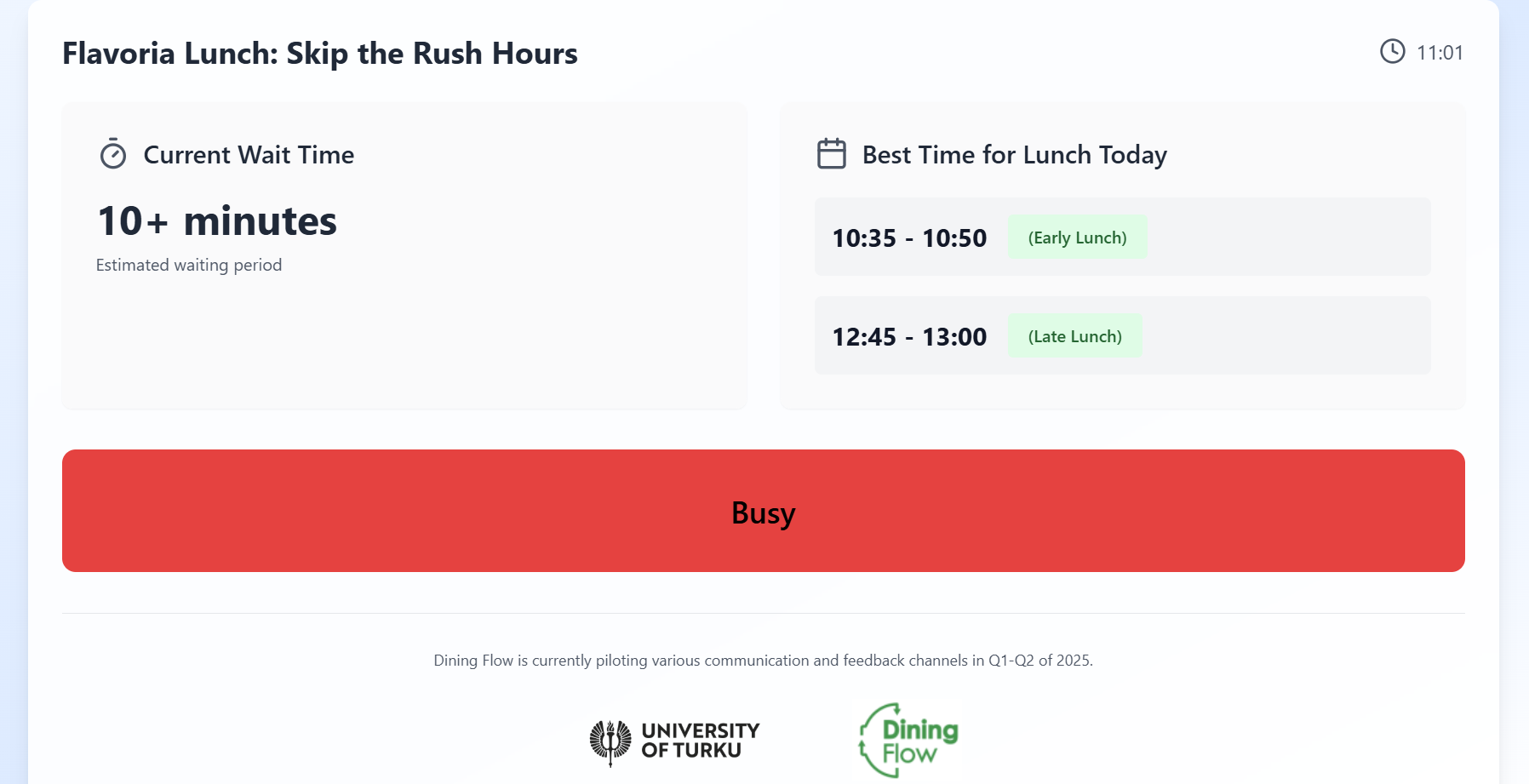}
    \caption{Passive Information Display: Busy State}
    \label{fig:passive-display-busy}
\end{figure}
\begin{figure}
    \centering
    \includegraphics[width=0.5\textwidth, alt={No wait times: Image showing there is No Wait and a green progress bar stating Quiet wait times}]{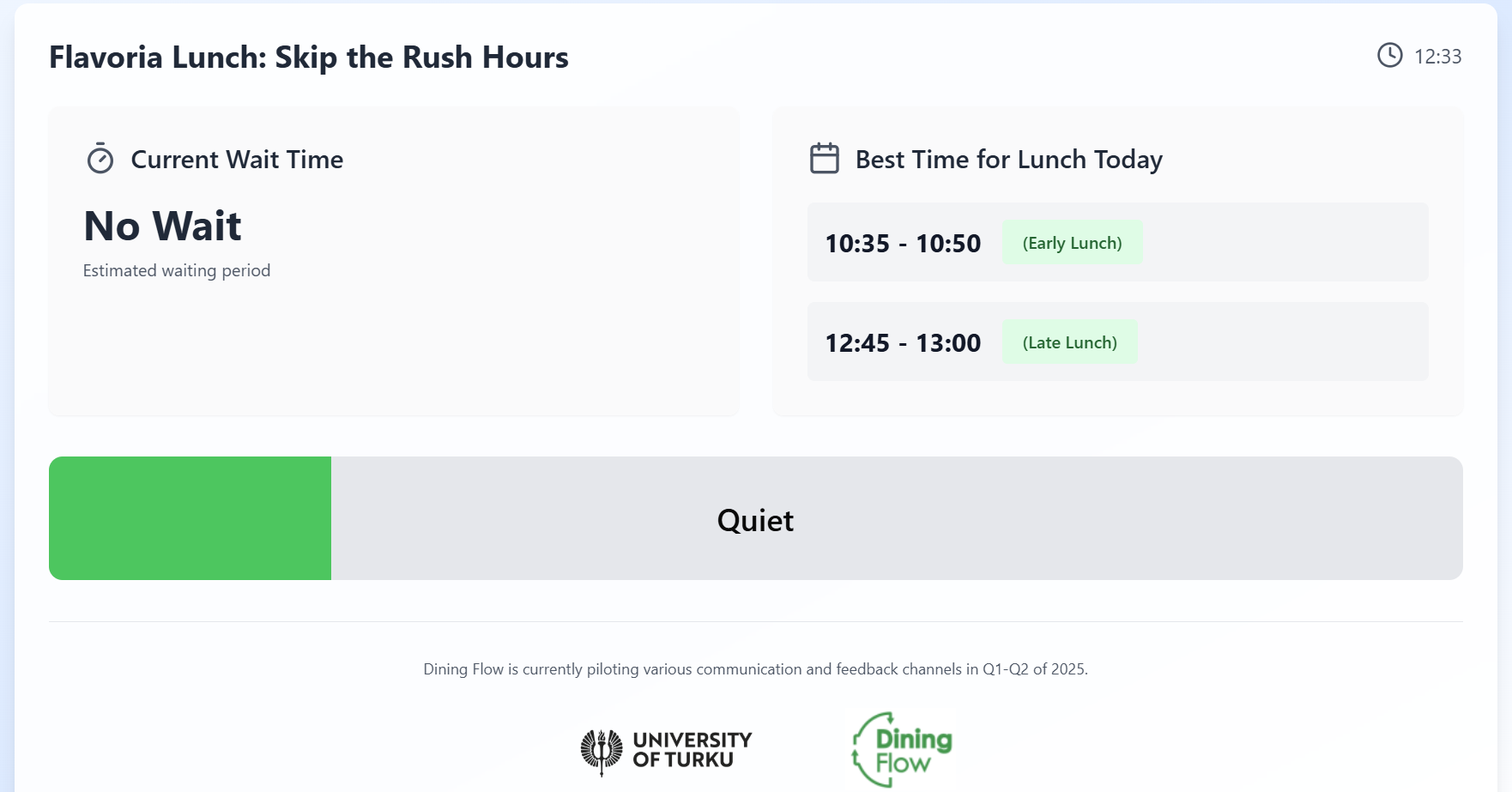}
    \caption{Passive Information Display: Quiet State}
    \label{fig:passive-display-quiet}
\end{figure}

Prior work in the same SSR~\cite{khursheed2025affordance} tested a passive "Rush Meter" displaying real-time occupancy levels and suggesting quieter dining windows with no interaction required. Table~\ref{tab:passive-display-states} and Figures~\ref{fig:passive-display-closed}, \ref{fig:passive-display-moderate}, \ref{fig:passive-display-busy} and \ref{fig:passive-display-quiet} show the different states of this display across the lunch period.

The Rush Meter had almost no impact on diners' behavior~\cite{khursheed2025affordance}. This showed that simply making information available on a screen is not enough and users need a reason to actively engage with the display. This raised a more fundamental question: when users do encounter an interactive screen, what determines the extent of their engagement, and how deeply? To explore this, we ran two experiments examining where users plateau in their engagement and whether prior commitment to an interaction draws them further in. Both experiments targeted the same audience, diners in the self-service restaurant, but used different interaction designs.

\begin{table}
 \centering
\tagpdfsetup{table/header-rows={1}}
  \begin{tabular}{ccll}
    \toprule
    Time & Status & Estimated Waiting Time & Reference Figure \\
    \midrule
    10:30--11:00 & Moderate & 5 minutes & Figure~\ref{fig:passive-display-moderate} \\
    11:00--12:15 & Busy & 10+ minutes & Figure~\ref{fig:passive-display-busy} \\
    12:15--12:30 & Moderate & 5 minutes & Figure~\ref{fig:passive-display-moderate} \\
    12:30--13:30 & Quiet & 0 minutes &  Figure~\ref{fig:passive-display-quiet}\\
    13:30--10:30 & Closed & N/A & Figure~\ref{fig:passive-display-closed} \\
    \bottomrule
  \end{tabular}
   \caption{Passive Information Display: Definition of different states}
     \label{tab:passive-display-states}
\end{table}

\subsection{Experiment Design}
Our analysis focuses on two experiments that examined user engagement with interactive displays requiring different levels of cognitive investment. These experiments provide insights into how interaction complexity influences engagement decisions under bounded rationality constraints. Two 88-inch interactive screens (one on entrance and one on exit) were used for each of the two experiments. 
In Experiment 1, the entrance display was oriented outward toward the corridor, so both diners and passers-by could see and interact with it. In Experiment 2, the display faced the dining hall and primarily targeted diners seated and having their lunch.\\
The source codes for the experiments in this study are available at the following URL: \url{https://github.com/ismailvohra/bounded-rationality-research-ssr/}.

\subsubsection{Experiment 1: Binary Choice Interface (Intentions Poll)} 
\begin{figure}
    \centering
    \includegraphics[width=0.5\textwidth, alt={Poll screen: Image showing a YES and NO button with the question Will you eat at Flavoria today}]{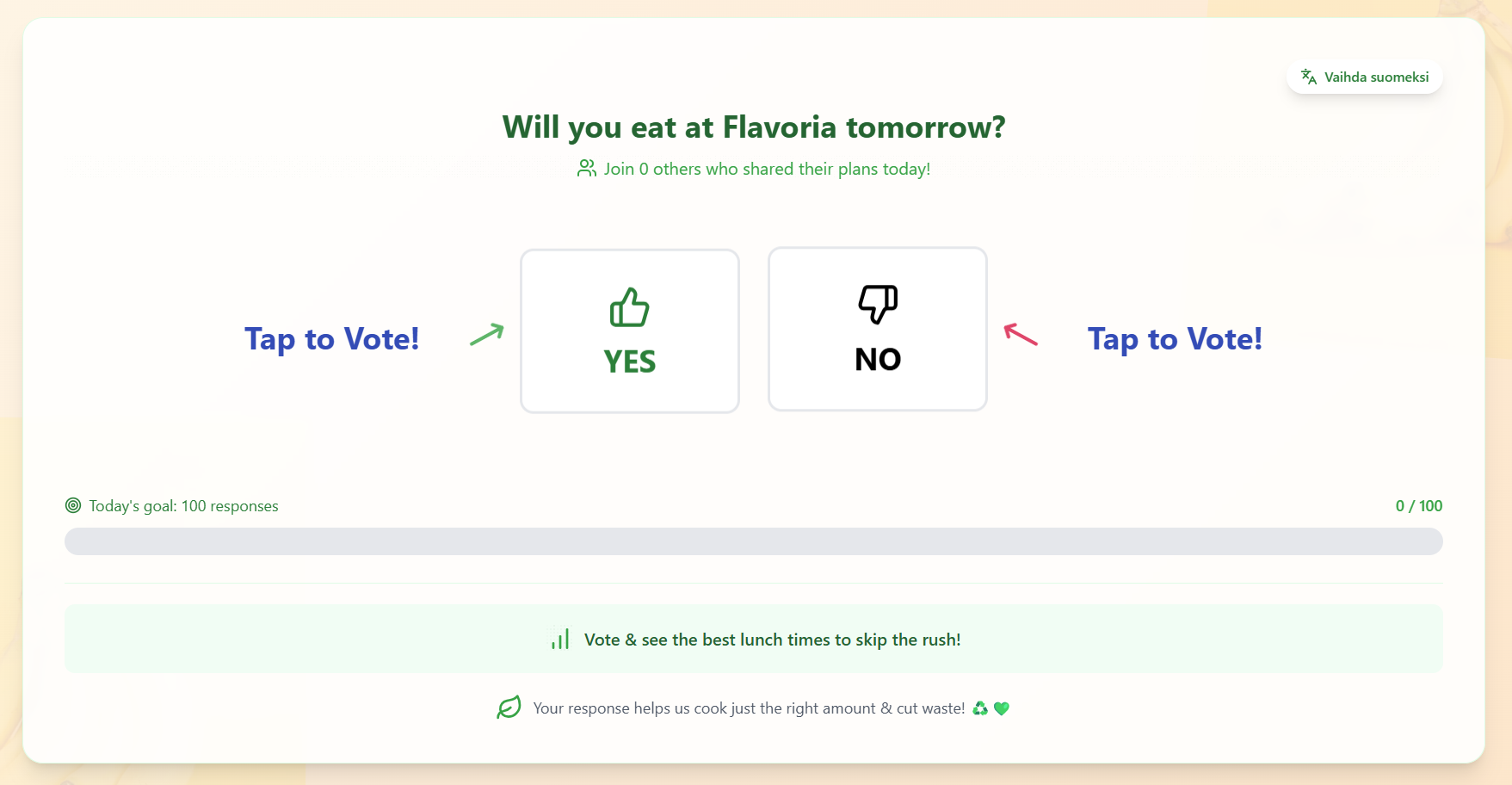}
    \caption{Snapshot of Experiment 1: Binary Choice Interface (Intentions Poll)}
    \label{fig:experiment-1}
\end{figure}

This experiment examined user engagement with a simple interface requiring minimal cognitive investment. Figure~\ref{fig:experiment-1} presents a screenshot of the screen design of this experiment.\\
The display presented a straightforward binary question: "Are you eating at Flavoria today?" with large, tappable YES and NO buttons. This design required explicit user action but minimal cognitive processing, representing low interaction complexity. The primary metric was the daily engagement rate calculated as the percentage of diners who responded "YES" to the question.

\subsubsection{Experiment 2: Multi-Step Interface (Food Waste Pledge)} \label{exp-2-design}
\begin{figure}
    \centering
    \includegraphics[width=0.5\textwidth, alt={Image of Tomato with Knife: Image showing a digital knife with animation to slice the tomato}]{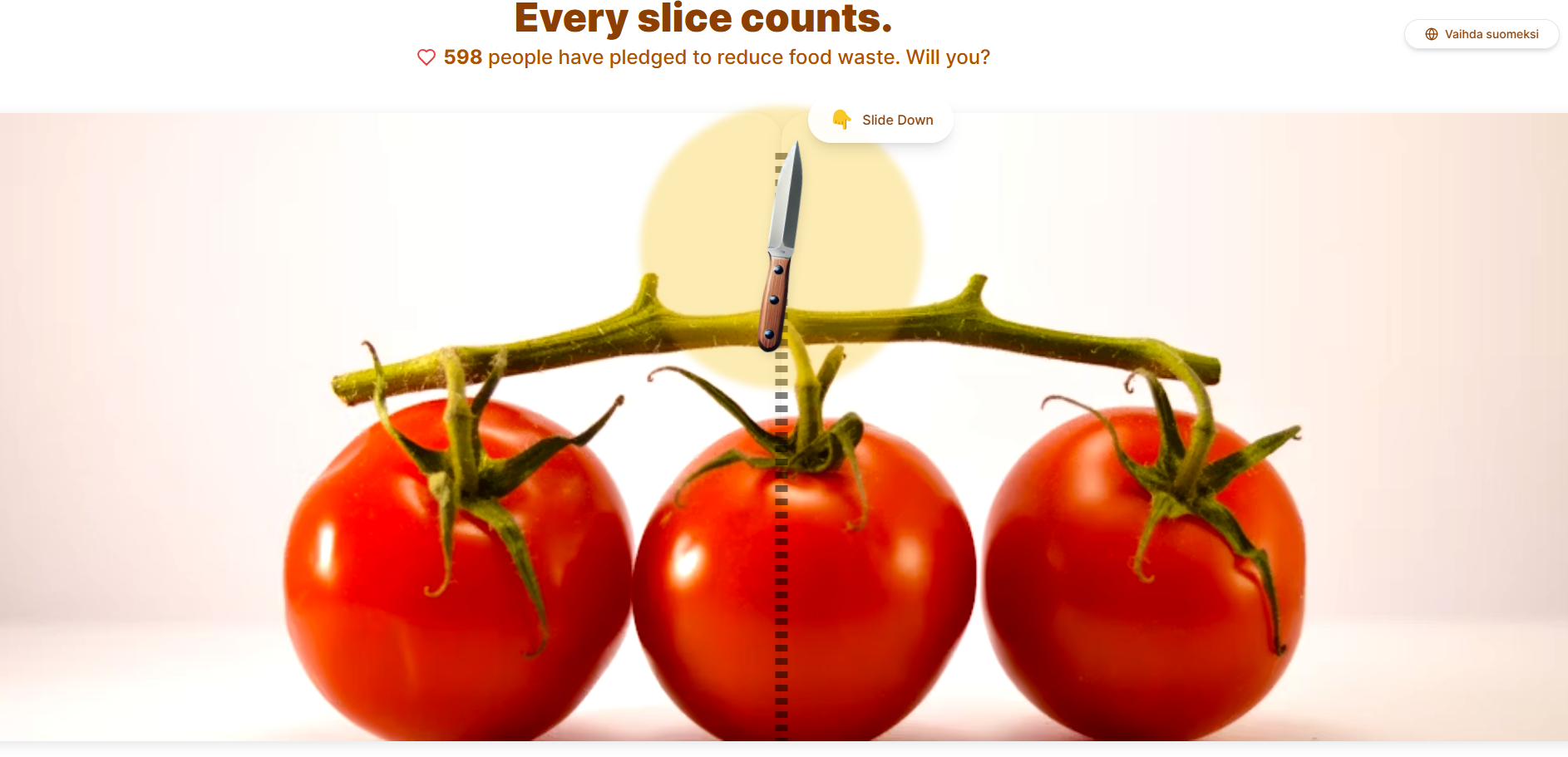}
    \caption{Snapshot of Experiment 2: Multi-Step Interface (Food Waste Pledge) - Slicing Tomato}
    \label{fig:experiment-2-tomato}
\end{figure}
\begin{figure}
    \centering
    \includegraphics[width=0.5\textwidth, alt={Image of Bread with Knife: Image showing a digital knife with animation to slice the bread}]{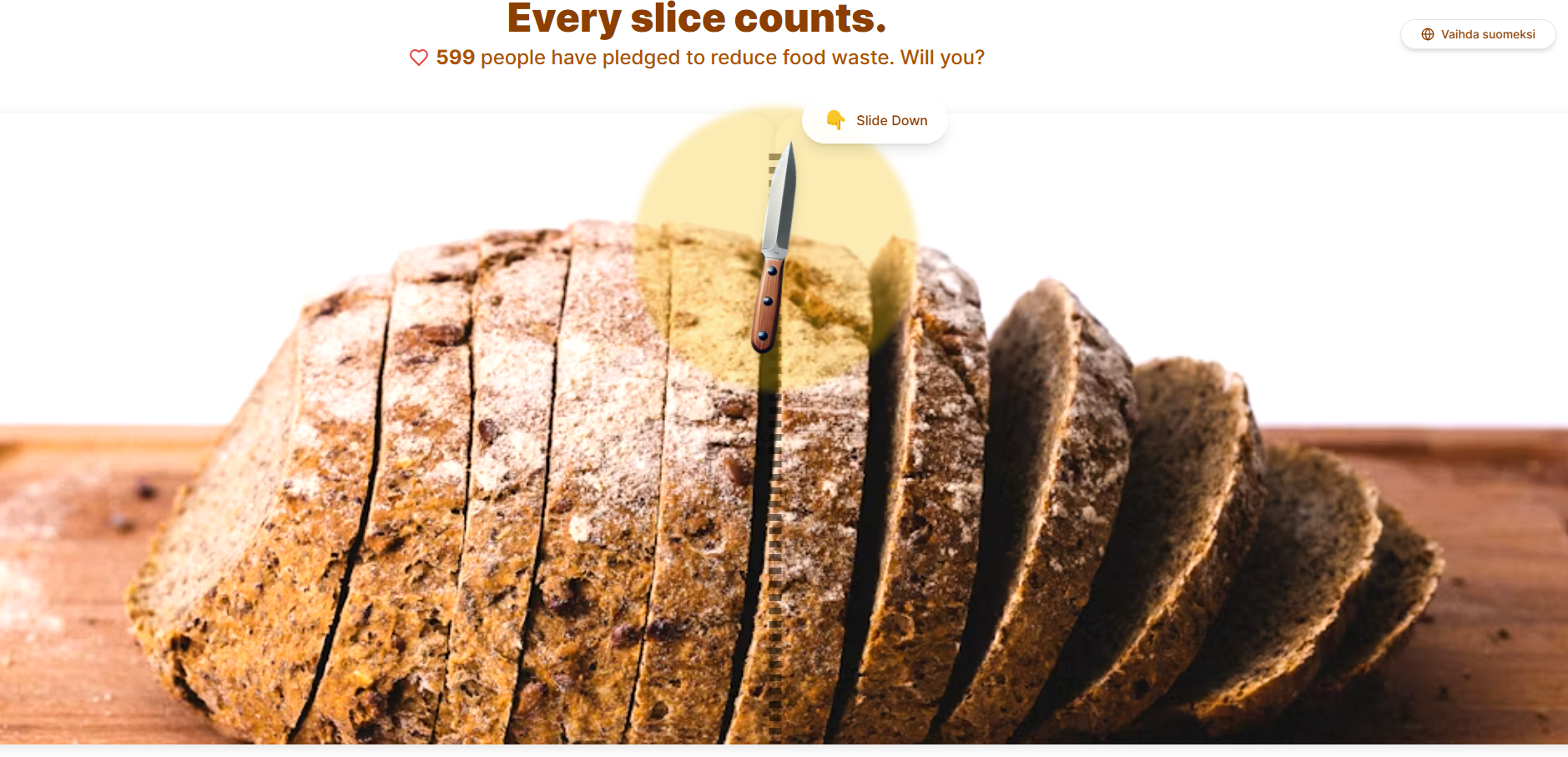}
    \caption{Snapshot of Experiment 2: Multi-Step Interface (Food Waste Pledge) - Slicing Bread}
    \label{fig:experiment-2-bread}
\end{figure}
\begin{figure}
    \centering
    \includegraphics[width=0.5\textwidth, alt={Image of Pizza with Knife: Image showing a digital knife with animation to slice the pizza}]{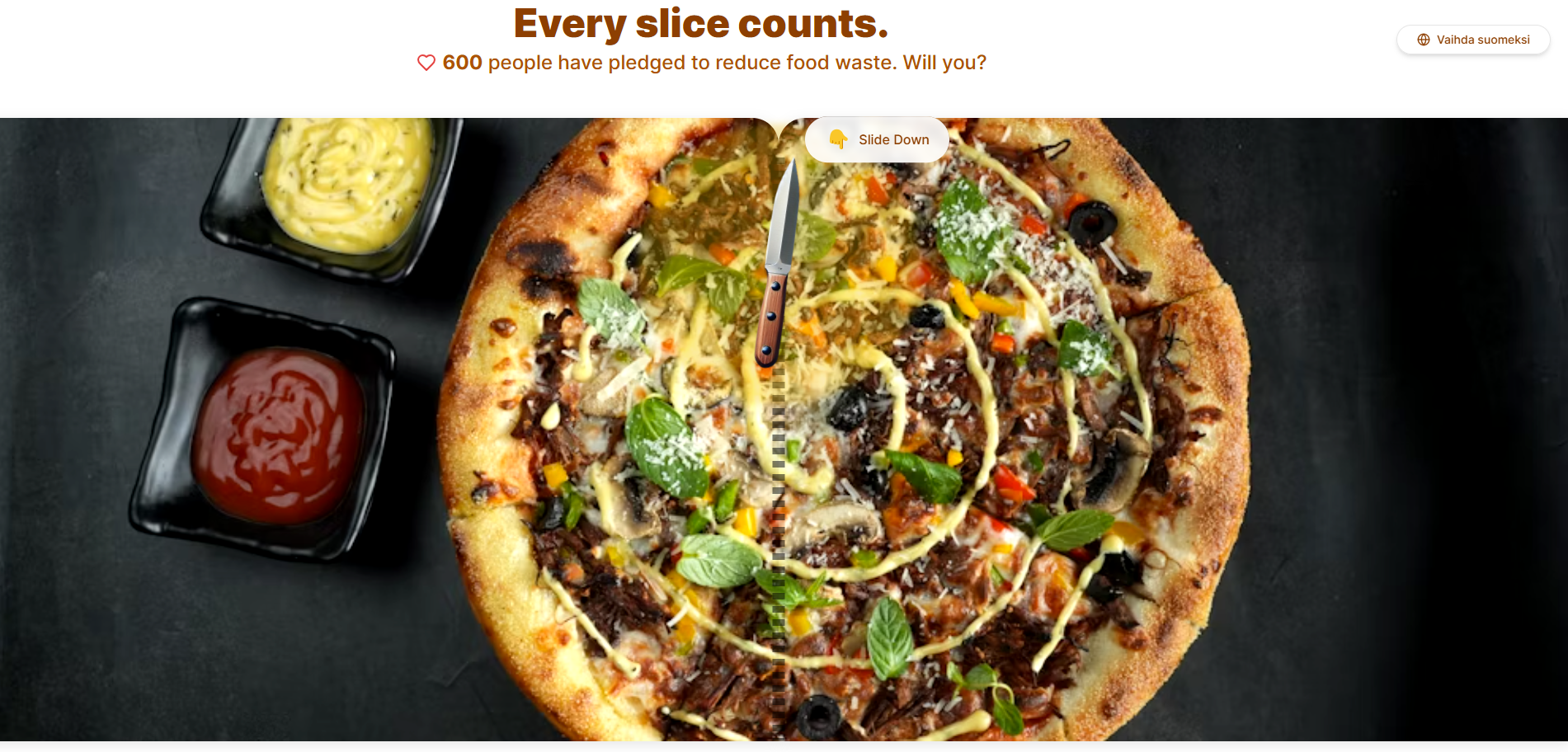}
    \caption{Snapshot of Experiment 2: Multi-Step Interface (Food Waste Pledge) - Slicing Pizza}
    \label{fig:experiment-2-pizza}
\end{figure}
\begin{figure}
    \centering
    \includegraphics[width=0.5\textwidth, alt={Take the pledge: Image showing a message of food waste situation in the restaurant and a button to take pledge to not waste food}]{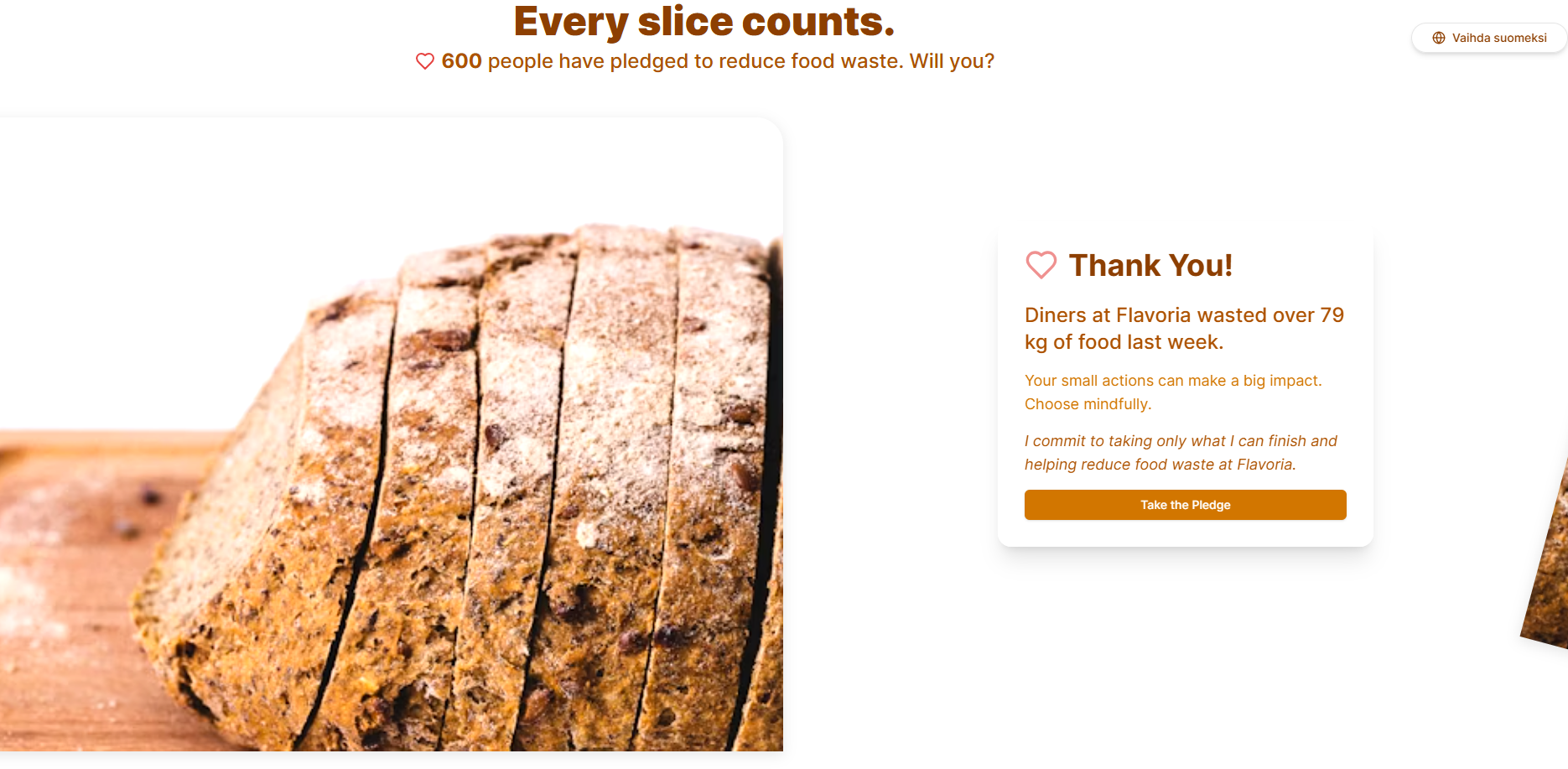}
    \caption{Snapshot of Experiment 2: Multi-Step Interface (Food Waste Pledge) - Pledge}
    \label{fig:experiment-2-pledge}
\end{figure}

This experiment examined user engagement with a complex interface requiring higher cognitive investment. Users were invited to perform a playful "slicing" gesture on digital food items using a virtual knife, followed by making a sustainability pledge to take only as much food as they could eat. This multi-step interaction represented high interaction complexity, requiring gesture recognition, decision-making, and commitment to behavioral change. The metrics included initial engagement rates and completion rates for users who started the interaction sequence.

Figures~\ref{fig:experiment-2-tomato}, \ref{fig:experiment-2-bread}, \ref{fig:experiment-2-pizza} and \ref{fig:experiment-2-pledge} present screenshots of the screen designs for this experiment. In this experiment, each diner saw a single image, and the display advanced to the next item in the sequence (tomato → bread → pizza → …) for the following diner.

\section{Results}

\subsection{Experiment 1: Binary Choice Interface (Intentions Poll)}
The binary choice interface experiment examined user engagement with a simple interface requiring minimal cognitive investment. The experiment ran from March 17th to March 28th, 2025, with daily interaction data logged automatically.

\begin{table}
\centering
\tagpdfsetup{table/header-rows={1}}
  \begin{tabular}{cccccc}
    \toprule
    Date & Diners & Interactions & Said YES & Said NO & \% YES to Diners \\
    \midrule
    17 Mar 2025 & 782 & 60 & 34 & 26 & 4.35\% \\
    18 Mar 2025 & 673 & 60 & 32 & 28 & 4.75\% \\
    19 Mar 2025 & 762 & 112 & 77 & 35 & 10.10\% \\
    20 Mar 2025 & 572 & 84 & 37 & 47 & 6.47\% \\
    21 Mar 2025 & 631 & 70 & 45 & 25 & 7.13\% \\
    24 Mar 2025 & 785 & 68 & 42 & 26 & 5.35\% \\
    25 Mar 2025 & 665 & 60 & 31 & 29 & 4.66\% \\
    26 Mar 2025 & 756 & 62 & 37 & 25 & 4.89\% \\
    27 Mar 2025 & 593 & 72 & 30 & 42 & 5.06\% \\
    28 Mar 2025 & 564 & 74 & 45 & 29 & 7.98\% \\
    \midrule
    Totals & 6783 & 722 & 410 & 312 & Avg: 6.04\% \\
    \bottomrule
  \end{tabular}
    \caption{Summary of Results for Experiment 1: Binary Choice Interface (Intentions Poll)}
  \label{tab:exp-1-results}
\end{table}

The average engagement rate was 6.04\% (410 YES responses out of 6,783 diners), with daily variation ranging from 4.35\% to 10.10\%. This variation suggests that contextual factors significantly influence user engagement decisions even with simple interactions.\\
Because the Experiment 1 screen also attracted passers-by whose total count is unknown, we normalize by the number of diners served that day, thus “\% YES to diners” is a per-diner-day rate, not the fraction of all exposed individuals who answered YES. \\
A summary of results for Experiment 1 is stated in Table~\ref{tab:exp-1-results}.

\subsection{Experiment 2: Multi-Step Interface (Food Waste Pledge)}
The multi-step interface experiment examined user engagement with a more complex interaction requiring higher cognitive investment. The experiment ran from April 14-28, 2025, with system logs tracking both initial interactions and pledge completions.

\begin{table}
\centering
\tagpdfsetup{table/header-rows={1}}
    \resizebox{\linewidth}{!}{%
  \begin{tabular}{cccccc}
    \toprule
    Date & Diners & 1st Interactions & 2nd Interactions (Pledge) & \% 1st Interactions to Diners & \% Completion Rate \\
    \midrule
    14 Apr 2025 & 635 & 54 & 37 & 8.50\% & 68.52\% \\
    15 Apr 2025 & 665 & 47 & 29 & 7.07\% & 61.70\% \\
    16 Apr 2025 & 657 & 37 & 22 & 5.63\% & 59.46\% \\
    17 Apr 2025 & 389 & 57 & 42 & 14.65\% & 73.68\% \\
    22 Apr 2025 & 646 & 43 & 17 & 6.66\% & 39.53\% \\
    23 Apr 2025 & 591 & 45 & 24 & 7.61\% & 53.33\% \\
    24 Apr 2025 & 771 & 37 & 16 & 4.80\% & 43.24\% \\
    25 Apr 2025 & 555 & 64 & 33 & 11.53\% & 51.56\% \\
    28 Apr 2025 & 632 & 31 & 10 & 4.91\% & 32.26\% \\
    \midrule
    Totals & 5541 & 415 & 230 & Avg: 7.49\% & Overall: 55.42\% \\
    \bottomrule
  \end{tabular}}
    \caption{Summary of Results for Experiment 2: Multi-Step Interface (Food Waste Pledge)}
  \label{tab:experiment-2-results}
\end{table}

The average initial interaction rate was 7.49\%, while the overall completion rate for users who started the interaction was 55.42\%. Daily completion rates varied significantly from 32.26\% to 73.68\%, indicating that environmental factors continue to influence user decisions even after initial engagement. \\
For comparability with Experiment 1, all percentages are normalized by diners served per day because this screen faced the dining hall. The denominator reflects diners rather than passers-by.\\
A summary of results for Experiment 2 is stated in Table~\ref{tab:experiment-2-results}.

\subsection{Cross-Experiment Comparison}
\begin{table}
\centering
\tagpdfsetup{table/header-rows={1}}
    \resizebox{\linewidth}{!}{%
  \begin{tabular}{ccccc}
    \toprule
    Experiment & Interface Type & Interaction Complexity & Engagement Rate & Completion Rate \\
    \midrule
    1: Intentions Poll & Binary Choice & Low (single binary choice) & 6.04\% & N/A (single step) \\
    2: Food Waste Pledge & Multi-Step & High (multi-step with gesture + commitment) & 7.49\% & 55.42\% \\
    \bottomrule
  \end{tabular}}
    \caption{Cross-Experiment Comparison Summary}
  \label{tab:experiments-summary}
\end{table}
The comparison reveals that despite significantly higher interaction complexity in the multi-step interface experiment, the engagement rate increased only marginally (from 6.04\% to 7.49\%), demonstrating an engagement plateau effect. However, users who overcame the initial barrier in the complex interaction demonstrated substantial commitment, with over half completing the full interaction sequence. Table~\ref{tab:experiments-summary} represents the cross-experiment comparison summary.

\section{Findings}
Our analysis suggests evidence of bounded rationality effects in user engagement with interactive displays in SSR environments. The data demonstrates how cognitive limitations and environmental constraints influence user decisions to engage with displays, providing insights into why users often stop at good enough rather than pursuing optimal interactions.

\subsection{Evidence for Engagement Plateau}
The comparison between binary choice and multi-step interfaces suggests evidence for an engagement plateau effect predicted by bounded rationality theory. Despite the multi-step interaction offering potentially greater value through meaningful sustainability engagement and incorporating playful elements, it achieved only a marginal increase in engagement rates (7.49\%) compared to the simpler binary choice interface (6.04\%).

This pattern aligns with bounded rationality predictions: when interaction complexity increases significantly, requiring gesture recognition, multi-step decision-making, and behavioral commitment, the cognitive cost creates barriers to initial engagement that offset potential benefits for most users. The minimal increase in engagement despite substantially higher complexity demonstrates that users have cognitive thresholds beyond which additional features become counterproductive.

The daily variation in engagement rates for both experiments (ranging from 4.35\% to 10.10\% for the binary choice interface and 4.80\% to 14.65\% for the multi-step interface) suggests that engagement thresholds are not fixed but fluctuate based on contextual factors such as crowd levels, time pressures, and social dynamics. This variability is consistent with bounded rationality theory, which predicts that decision-making strategies adapt to current constraints and available cognitive resources.

\subsection{High Completion Rates: The Commitment Effect}
The 55.42\% completion rate for the multi-step interface experiment reveals insights into how bounded rationality affects user behavior after initial engagement. This relatively high completion rate, despite the interaction's complexity, suggests that users who overcome the initial cognitive barrier to engagement are likely to continue through multi-step interactions.

This pattern can be explained through bounded rationality mechanisms. Once users invest cognitive effort in starting an interaction, the sunk cost of attention and mental effort already invested makes completion more attractive than abandonment. Additionally, users who choose to engage with complex interactions demonstrate both higher motivation and available cognitive resources, making them more likely to persist through completion.

The significant variation in daily completion rates (32.26\% to 73.68\%) indicates that environmental factors continue to influence user decisions even after initial engagement. Users appear to continuously evaluate whether the ongoing cognitive cost justifies completing the interaction, with contextual factors such as time pressure, social dynamics, and competing priorities affecting this calculation.

\section{Discussion}
Our findings suggest three design principles for interactive displays in constrained environments:

Minimize Initial Cognitive Barriers: The engagement plateau demonstrates that initial interaction complexity significantly affects user willingness to engage. Designs should prioritize simple, immediately comprehensible entry points that require minimal cognitive processing. Complex functionality can be introduced progressively for users who demonstrate initial engagement. 

Leverage Progressive Engagement: The high completion rates for users who start complex interactions suggest that progressive engagement strategies can be effective. Designs can use simple initial interactions to identify motivated users, then provide opportunities for deeper engagement for those willing to invest additional cognitive effort.

Balance Complexity and Value: While increased complexity reduces initial engagement, meaningful interactions can still attract users willing to invest cognitive effort. Designs should carefully balance interaction complexity against the perceived value and personal relevance of the interaction outcomes.

\section{Limitations}
Several limitations should be acknowledged in this research:

Single-site study: All experiments were conducted at one university SSR, which may limit the applicability of findings to other public environments with different user populations, cultural contexts, or operational characteristics.

Limited experimental scope: This analysis focuses on two experiments examining specific interaction modalities. Future research incorporating varying complexity levels and interaction types would strengthen understanding of engagement plateau effects and bounded rationality patterns.

Temporal constraints: The experiments were conducted over specific time periods that may not capture seasonal variations, long-term shifts in user patterns, or adaptation effects that could influence user engagement patterns.

Demographic analysis: While the user population was diverse, detailed demographic analysis was not conducted, preventing examination of how age, technological literacy, or cultural background might influence engagement decisions under bounded rationality constraints.

\section{Future Work}
Future work could explore several promising directions. First, extending the experimental program to investigate bounded rationality effects across different interaction modalities would strengthen the empirical foundation for engagement plateau concept and deepen understanding of cognitive thresholds in public display contexts. Replicating these findings across different public environments would also help establish how generalizable bounded rationality patterns are beyond the specific setting of self-service restaurants. Longitudinal studies would further complement this by revealing how user engagement evolves over time, including adaptation effects, learning curves, and long-term behavioral changes in response to interactive displays.

\section{Conclusion}
This paper investigated user engagement patterns with interactive displays of varying complexity in self-service restaurant environments through the lens of bounded rationality theory. Our analysis of two field experiments at Flavoria Research Lab suggests empirical evidence for how cognitive limitations and environmental constraints influence user decisions to engage with public interactive displays.

The identification of engagement plateaus provides a theoretical lens for understanding how interaction complexity affects user behavior in constrained SSR  environments. This concept can inform design decisions about when to prioritize simplicity versus functionality, and how to structure progressive engagement strategies that respect human cognitive limitations.

Our initial findings for completion patterns in multi-step interactions contributes to understanding how initial cognitive investment influences subsequent user behavior. The finding that users who overcome initial barriers demonstrate high commitment to completion provides opportunities for designing more sophisticated interactions for motivated users while maintaining simple entry points for broader accessibility.

This work demonstrates that understanding human cognitive limitations supports the effective designing of public interactive systems. By recognizing and accommodating bounded rationality, designers can create interfaces that work with natural human decision-making processes, leading to more successful user engagement in constrained environments. The theoretical lens and empirical findings presented here provide a foundation for applying cognitive theories in practice, advancing our understanding of people, guiding future research in human–computer interaction in public spaces, and informing the design of systems that respect human cognitive abilities.

\section*{Acknowledgments}
This research was supported by Business Finland, under the Veturi program with the Dining Flow project (6547/31/2022).

\bibliographystyle{unsrtnat}
\bibliography{references}

\end{document}